# Deformation of Heartbeat Pulse Waveform Caused by Sensor Binding Force

Jiesheng He[1*] and Wei Wu[2*]

[1] *Canyon Crest Academy, San Diego, CA, USA*
[2] *Dept. of Electronical and computer Engineering, USC, Los Angeles, CA, USA*
*Email: jhjasonhe@gmail.com, wu.w@usc.edu

*Abstract*--**Photoplethysmography (PPG) is a method of detecting variation in blood volume commonly through contact with the skin and involving the usage of one or multiple sensors. PPG is typically used in health-related fields and one of its most popular uses is heart rate monitoring. Recent developments in technology have introduced wearable smart devices that can detect PPG signals. These devices mostly rely on signals that are detected at a single location, indicating that such devices require the signal waveform to be immensely reliable. In further detail, the properties of the blood circulation system are being deducted from the heartbeat signal profile. One disadvantage is that the waveform can be affected by tissues that surround the blood vessels. This may result in the unreliability of the methods utilized by many wearable smart devices. This paper introduces an experimental study on the deformation of waveform change caused by sensor binding force. The results of the study are that an increase in sensor binding force decreases the width of the pump strength-related pulse.**

## I. Introduction

Cardiovascular diseases (CVDs) are a serious health concern in many countries around the world. In 2019, the World Health Organization (WHO) estimated that 17.9 million people, a number that made up approximately 32% of the total global deaths that year, died due to complications caused by CVD [1]. While the majority of deaths caused by CVDs occur in countries that are classified as lower or middle income, high-income nations do not appear to be immune: CVDs are among the leading causes of death in the United States [2], Germany [3], France [4], South Korea [5], and Spain [6].

Despite their relatively high mortality rate, CVDs are treatable. Treatments have previously shown to have great influence in decreasing CVD mortality rates. Japan, for example, has attributed 56% of its 61% decrease in coronary heart disease mortality rates between 1980 and 2012 to surgical and medical treatments [7]. Research is currently being conducted on a number of potential new CVD treatment methods including 3D printing, nanotechnology, robotic surgery, drugs, and stem cells [8]. Most of the aforementioned methods have already shown promise and researchers believe they will ease the process of treating CVDs in the future.

CVDs are also preventable. The Centers for Disease Control and Prevention (CDC) has listed several ways individuals can prevent themselves from being affected by CVDs, one of them being regularly monitoring and controlling blood pressure [9]. Previously, this required people to visit a clinic or hospital. Thanks to recent advances in technology, however, people can now monitor their blood pressure from a wearable smart device like a watch anytime and anywhere they wish. This option not only offers flexibility, but it also eliminates the cost of clinics or hospitals. Due to these advantages, wearable smart devices have seen an increase in popularity in recent years.

One method that is being used by wearable smart devices to monitor blood pressure is photoplethysmography (PPG). PPG essentially measures how much light is being absorbed or reflected by blood vessels [10]. PPG can directly detect and measure changes in blood volume since the amount of blood present influences the amount of light being absorbed or reflected [10]. While PPG signals do not directly measure nor detect blood pressure, they can still be used to indirectly estimate blood pressure via several specialized techniques [11]. PPG also has many benefits, including its low cost, versatility, and simplicity.

However, PPG also comes with several disadvantages. The main drawback is the variety of factors that can affect the accuracy of PPG. As PPG relies on one or multiple sensors, it is unsuitable in conditions where unrestricted movement is required [12, 13]. This is mainly due to the high sensitivity of PPG signals to movements, which ultimately limits the scope in which PPG can be used reliably [12]. Other factors that have been known to affect PPG are skin damage [12], temperature [13], sensor geometry [14], light intensity [14], and oxygen concentration [14].

Most importantly, PPG can also be affected by an individual's internal organs, especially the tissues surrounding the blood vessels. This can severely affect the accuracy and reliability of the PPG signals that wearable smart devices use to measure blood pressure. In addition, it has also been noticed that current wearable smart devices also lack clinical accuracy when analyzing time components such as the pulse transit time. We hypothesized that these inaccuracies may have resulted from differences in the pressure being applied to the arm by a device such as a cuff or the touch force being applied by the finger to the PPG sensor. We devised an experiment to test our hypotheses and to find solutions that one can employ to minimize any inaccuracies. Our findings could be used to improve the accuracy of the PPG method that wearable smart devices use to monitor blood pressure. This could potentially also aid in the prevention of CVDs.

## II. Experiment Setup

As shown in Figure 1, we used the heartbeat sensor of a Samsung Galaxy Note 9 phablet to capture the pulse waveforms. While this is not exactly identical to wearable smart devices, it is similar enough to carry out the experiment without any major consequences. We used a scale, manufactured by AmazonBasics, to measure the touch force in grams. This will be used to check any pulse waveform changes with a locally binding force.

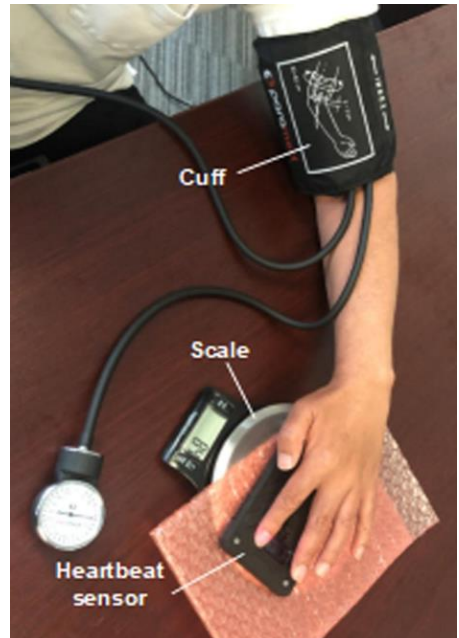

Figure 1: Heartbeat pulse waveform sensing at different touching forces. Note that the actual

blood pressure is not changing. The experiment is entirely a simulation.

The PPG method sometimes utilizes a multiple-sensor setup instead of a single-sensor setup. These sensors would be mounted at different positions. In our study, a cuff was mounted at the upper arm to apply varying amounts of pressure. The variation of the pressure simulates the sensor binding force change. It should be noted that most commercial wearable smart devices like watches are cuffless. The force applied by the finger touching the sensor would remain the same when experimenting with different pressures. The heartbeat pulse waveform is monitored in order to note the influence from the pressure being added by the cuff. When testing for influences from the touch force, the pressure was kept at a constant amount and varying finger touch forces were applied.

We also investigated regular commercial heartbeat sensors. We determined that the heartbeat pulse waveform is usually generated by the circuit inside the sensor. The heartbeat signals would trigger the circuit to send out the pulses and the waveform, with the exception being the amplitude, would remain similar. Our method of study modifies the blood pressure through changes in blood flow and resistance. More than ten trials were performed for each variable and the results were able to be repeated multiple times. This has the potential to increase the reliability of the experiment.

### III. About Heartbeat Waveforms

Blood volume refers to the amount of blood that is present in the body at any given time. The blood volume is influenced by an individual's size and age, among other factors. Optical heartbeat sensor signals are created by the blood flow volume change. A large volume will result in more absorption. It is important to note that the sensor output will occasionally show the result upside-down.

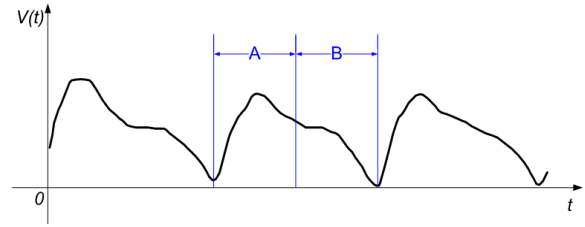

Figure 2: Example of a heartbeat pulse waveform. Halves *A* and *B* are explicitly marked.

As shown in Figure 2, a heartbeat cycle can be roughly divided into two halves: *A* and *B*. *A* represents heart pump conditions like the volume and pump strength, while *B* reflects the resistance or recoil of the blood vessels. The pump strength refers to how well the heart is able to pump blood. The heart has left and right ventricles. The left ventricles are responsible for pumping oxygenated blood to the rest of the body while the right ventricles pump deoxygenated blood to the lung for oxygenation. Meanwhile, the atria receive the blood and deliver it to the ventricles. The right atrium receives deoxygenated blood while the left atrium receives oxygenated blood. The resistance refers to the blood vessels opposing the blood flow. This depends on the diameter of the blood vessel as smaller diameters cause greater resistance and decreased blood flow.

Arteries are blood vessels that consist of three layers: intima, adventitia, and media. The main function of arteries is carrying oxygen-rich blood from the heart to tissues throughout the body. The one exception to this are the pulmonary arteries that carry blood from the heart to the lungs. This helps to oxygenate the blood. Veins are blood vessels that are thinner and relatively weaker than arteries. The primary function of the veins is to carry blood back to the heart. The veins are connected to the arteries by capillaries. In addition, the blood flowing through veins has a lower pressure than blood flowing through arteries. As a result, veins are located closer to the skin surface than arteries. This also means that veins will be affected before arteries when pressure is being applied to the arm and its tissues. Additionally, this also indicates that the blood flow resistance will be changed first.

### IV. Waveform Change by Cuff Pressure

Figure 3 shows heartbeat pulse waveforms that were generated at the left index fingertip. The touch force was kept constant at approximately 30 g as weighed by the scale and the initial upper arm blood pressure was measured to be 119/78 mmHg. The heart rate is 64bpm.

It is apparent that the waveforms change when the cuff pressure is increased. At a cuff pressure of 0 mmHg (when the cuff is not applying additional pressure), the heartbeat waveform indicates the lowest blood flow resistance. When the cuff pressure is increased to 60 mmHg, a little below the diastolic threshold, the pulse profile appears to sink at the right half. When the cuff pressure is increased to 80 mmHg, which is around the diastolic threshold, the right side of the curve becomes flat while the peak on the left side looks narrower. When the cuff pressure is further increased to 90 mmHg and 100 mmHg, the peak on the left side of the curve becomes increasingly narrow. This trend continues as the pressure approaches the systolic threshold where only a miniscule amount of the peak on the left side can be seen.

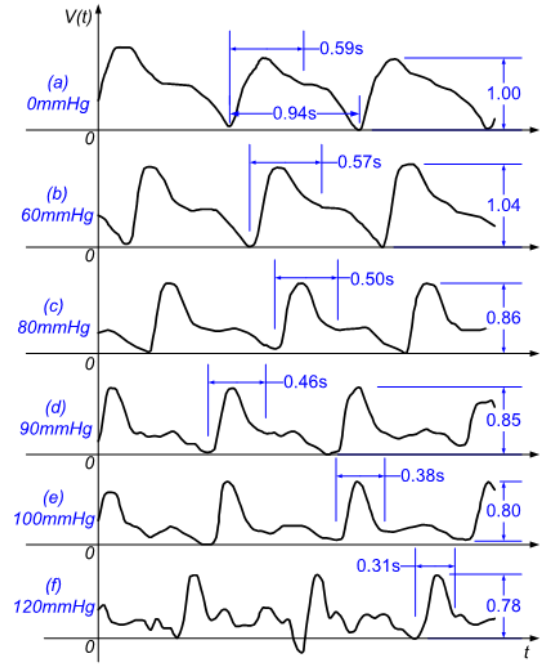

Figure 3: Heartbeat pulse waveforms (a-f) with varying cuff pressures of 0 mmHg, 60 mmHg, 80 mmHg, 90 mmHg, 100 mmHg, and 120 mmHg. The initial upper arm blood pressure is 119/78 mmHg and the touch force is kept constant at approximately 30g. The amplitudes (a.u.) are relative to the first waveform. The time interval (s) of the heart pump strength-related pulse is marked for each waveform.

Thus, the pulse signal amplitude generally decreases while the cuff pressure increases. This means that the relative blood volume change ratio is being reduced. In addition, the time interval of the heart pump strength-related pulse decreases as the added cuff pressure increases. The amplitude of the waveforms also appears to decrease as the added cuff pressure increases, with the increase of amplitude between 0mmHg and 60mmHg being a notable exception to this trend.

It is important to note that the pulse waveform sensor and the cuff are mounted at different positions. Based on the waveforms observed in this experiment, it is rather difficult to trace stable parameters such as the PAT or PTT, which are crucial for the PPG method.

## V. Waveform Change by Touch Force

Figure 4 shows diagrams of heartbeat pulse waveforms that were generated at the left index fingertip. It is apparent that the waveforms change when the touch force is increased.

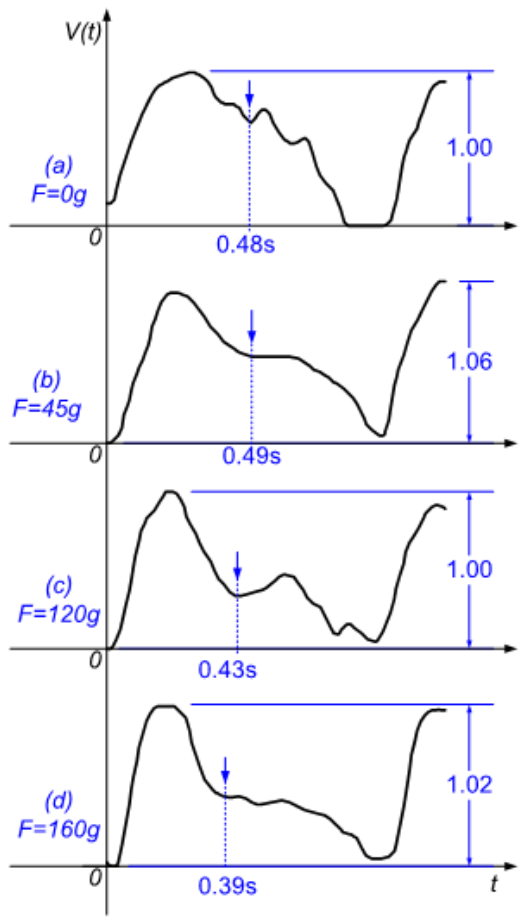

Figure 4: Heartbeat pulse waveforms (a-d) with varying touch forces (0g, 45g, 120g, and 160g). The systolic blood pressure roughly corresponds with 200 grams of press/touch force. The upper arm blood pressure is 119/78 mmHg. The amplitudes (a.u.) are relative to the first waveform. The time (s) where the first trough appears is marked for each waveform.

When the fingertip does not touch the sensor surface (which corresponds to a touch force of 0 g), the heartbeat waveform reveals the lowest blood flow resistance. When the fingertip presses against the surface of the sensor with a force of 45 g, the pulse profile sinks at the right half. When the fingertip touch force is increased to roughly 120 g, the sinking near the center of the waveform becomes more apparent. We can also see a second peak appearing. When the touch force is increased to about 160 g, the second peak on the right disappears. Thus, based on the waveforms that were observed, it is rather difficult to obtain stable parameters like PAT and PTT which are crucial for the PPG method.

Additionally, when the touch force increases, the time it takes for the first trough in the waveform to appear decreases. The increase from 48 seconds to 49 seconds between 0g and 45g of touch force is the sole exception. There does not seem to be a notable trend in the peak amplitudes of the waveforms.

## VI. Conclusion

The heartbeat pulse waveform can be easily affected by external forces. Both cuff pressure and contact force appear to affect the heartbeat pulse waveform. This is because both increasing cuff pressure and increasing contact force influence the blood vessels and their surrounding tissues. As the cuff pressure and contact force increases, the changes to the waveform increase as well and become increasingly apparent. This potentially can compromise the accuracy and reliability of the PPG method that many wearable smart devices employ to measure blood pressure. To minimize the effects, we strongly recommend that the design of the sensor installation be completed extremely carefully in the applications that require pulse waveform details. For example, this can be accomplished by keeping binding force constant with a structure that can automatically adjust its installation belt length.